\documentclass[aps,prb,twocolumn,groupedaddress,showpacs]{revtex4}
\usepackage{epsfig}
\usepackage{color}
\begin{document}

\title{
Determinant Quantum Monte Carlo Study of the
Orbitally Selective Mott Transition
}
\author{K. Bouadim$^1$, G.G. Batrouni$^2$, and R.T. Scalettar$^3$} 
\affiliation{
$^1$Physics Department, Ohio State University,
Columbus, OH 43202
}
\affiliation{
$^2$INLN, Universit\'e de Nice-Sophia Antipolis, CNRS; 
1361 route des Lucioles, 06560 Valbonne, France
}
\affiliation{
$^3$Physics Department, University of California, Davis, CA 95616
}

\begin{abstract}
We study the conductivity, density of states, and magnetic
correlations of a two dimensional, two band fermion Hubbard model
using determinant Quantum Monte Carlo (DQMC) simulations.  We show
that an orbitally selective Mott transition (OSMT) occurs in which the
more weakly interacting band can be metallic despite complete
localization of the strongly interacting band.  The DQMC method allows
us to test the validity of the use of a momentum independent
self-energy which has been a central approximation in previous OSMT
studies.  In addition, we show that long range antiferromagnetic
order (LRAFO) is established in the insulating phase, similar to the
single band, square lattice Hubbard Hamiltonian.  Because the critical
interaction strengths for the onset of insulating behavior are much
less than the bandwidth of the itinerant orbital, we suggest that the
development of LRAFO plays a key role in the transitions.
\end{abstract}

\pacs{
}

\maketitle

\noindent
\underbar{Introduction:} The problem of a strongly correlated band put
in contact with a more weakly interacting one is of long-standing
interest.  In the case of the periodic Anderson model (PAM), for
example, one orbital is completely free of interactions, while a second
orbital is at the opposite extreme: it has no hopping from site-to-site
(zero bandwidth) and instead has only an on-site hybridization $V$ with
the uncorrelated band.  A competition between on-site singlet formation
between electrons in the two different orbitals and RKKY mediated
antiferromagnetic (AF) order occurs as a function of $V$, and a
resonance in the density of states at the Fermi surface is present at
the transition between these two regimes.

Recently there have been a number of studies,
\cite{liebsch04,arita05,inaba06,orbselective} mainly within dynamical
mean field theory (DMFT) \cite{DMFT},
of the general question whether two
different bands can exist with one metallic and the other insulating,
the so-called `orbitally selective Mott transition' (OSMT).  Alternate
methods of treating the correlations of the impurity problem arising
within DMFT, ranging from iterated perturbation theory
\cite{liebsch04} to Quantum Monte Carlo (QMC) \cite{liebsch04,arita05} and exact
diagonalization \cite{liebsch05,inaba06} yield different results.  The form of
the interband coupling and, specifically, whether the Hund's rule term
is treated in an SU(2) symmetric way or only an Ising term is
retained, was also thought to affect the results.  By now it is
established that, within DMFT and using the most accurate impurity
solvers, an OSMT is possible.  As might be expected, the narrow band
becomes insulating first, as correlations increase, followed by the
wide band.  Attention has also focussed on the nature of the
transitions which are, in general, believed to be first order.

The most well controlled theoretical work on the OSMT has been
formulated within the framework of model Hamiltonians (multiband
Hubbard models) whose simplicity allows for detailed and precise
numerical studies.  However, similar issues have also been addressed
using a combination of electronic structure and many-body methods to
describe real materials.  The Cerium volume collapse
transition\cite{allen82} is one prominent example in which there is
an interplay between the localized $f$ and metallic $d$ orbitals.  As
in the PAM, the orbitals see each other through hopping processes as
well as interaction.  Here, Kondo physics arising from singlet
formation between electrons in the two bands plays a crucial role in
the volume collapse transition\cite{allen82,held01}.  Similar OSMT physics
occurs in other materials, Ca$_{2-x}$Sr$_x$RuO$_4$ being especially
well studied\cite{CaSrRuO}.

The range of different conclusions which arise depending on the
treatment of the many body correlations within DMFT suggests that
there is a crucial need to examine also the role of the local DMFT
approximation itself.  In this paper we use determinant Quantum Monte
Carlo (DQMC) to study the OSMT.  This method allows us to test
rigorously the effect of ignoring momentum dependence in the
self-energy within DMFT as well as to examine the real space AF
correlations which could form along with the Mott insulating
transition.

\noindent
\underbar{Model and Computational Method:}
We consider a Hamiltonian in which there is one correlated electron
band and a second orbital which is fully localized and represented by
a set of spin-$\frac{1}{2}$ degrees of freedom,
\begin{eqnarray} 
\label{hubham} 
H &=& -t \sum_{\langle ij \rangle \sigma} 
( \, c^\dagger_{i\sigma} c_{j\sigma}
 + c^\dagger_{j\sigma} c_{i\sigma} \, )
 - \mu \sum_i (n_{i\uparrow} + n_{i\downarrow} )
\nonumber \\
 &+& \sum_i  \Big[ \, 
 J_z S^z_i (n_{i\uparrow} - n_{i\downarrow})
+ J_{\perp} ( \, S^+_i c^\dagger_{i\downarrow} c_{i\uparrow}
 +  S^-_i c^\dagger_{i\uparrow} c_{i\downarrow} \, ) \,
 \Big] 
\nonumber \\
 &+& U \sum_i (n_{i\uparrow} - \frac{1}{2}) (n_{i\downarrow} - \frac{1}{2})
\end{eqnarray} 
Here $t$ allows the hopping of electrons of spin $\sigma$ between
adjacent sites $\langle ij \rangle$ of a square lattice, with
$c_{i\sigma}^\dagger(c_{i\sigma},n_{i\sigma})$ the associated creation
(destruction, number) operators.  $U$ is the on-site repulsion.  We
chose $t=1$ as our energy scale.  A chemical potential $\mu$
controls the filling.  We set $\mu=0$ which, by particle-hole
symmetry, pins the density at half-filling, $\rho=1$.  These fermions
are coupled to a set of local spin-$\frac{1}{2}$ degrees of freedom 
$S_i$ at each lattice site.  In this paper we will restrict the Hund's
rule interaction to the $J_z$ term, as has been done by Costi {\it et
al} in a recent DMFT study\cite{costi07}.

In the Hamiltonian Eq.~1 the local spins represent a highly localized
orbital; hence the question of the possibility of an OSMT devolves to
whether a metal-insulator phase change can occur in the remaining
itinerant fermion orbital as the energy scales $J$ and $U$ are tuned.
Eq.~1 is closely related to the Kondo lattice model, except that an
on-site $U$ is present for the electronic degrees of freedom, which is
usually set to zero in the Kondo case.  Changing $U$ allows,
potentially, for tuning through an OSMT.  A number of experimental
systems can be approximately described by such a mixture of electrons
and spins \cite{imada98}.  There are other materials whose qualitative
physics has been suggested to be described by Eq.~1, including
Ca$_{2-x}$Sr$_x$RuO$_4$ where a spin-$\frac{1}{2}$ Ru ion moment
coexists\cite{costi07,nakatsuji0300,liebsch06} with a metallic state near
$x=\frac{1}{2}$.

Our methodology is a version of the DQMC\cite{blankenbecler81}
algorithm often used to study Hubbard Hamiltonians, modified to
include the effects of the fluctuating local spin degrees of freedom
which represent the localized band.  These local spins, together with
the Hubbard-Stratonovich (HS) field which decouples the interaction,
specify the up and down spin determinants whose product acts as the
weight for the combined HS and local spin configuration.  The HS field
depends on both the spatial site and on the imaginary time coordinate
$\tau$ which arises when the inverse temperature $\beta$ is discretized.
The local spin, on the other hand, while varying in space, is constant
in $\tau$.  The HS variables are updated with the usual fast algorithm
which uses the fermion Green's function to compute the change in the
determinant \cite{blankenbecler81}.  The local spin is updated with a
variant of the approach used for `global moves' to ensure ergodicity in
the HS distribution in determinant QMC \cite{scalettar91}, since those
moves were also developed to handle changes which are non-local in
$\tau$.  

\begin{figure}[t]
\epsfig{figure=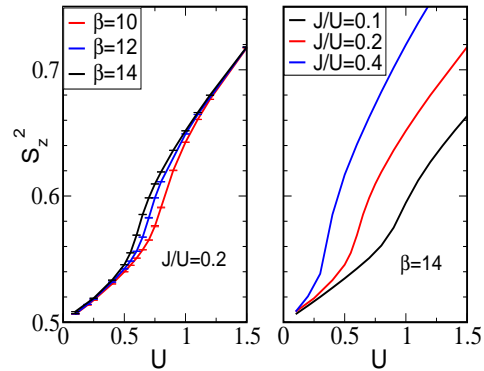,height=7cm,width=6cm,angle=-90,clip}
\caption{(Color online) Left: The local moment $\langle S_z^2 \rangle$
  of the correlated electron band is shown as a function of $U$ for
  three different values of inverse temperature $\beta$ for fixed
  $J/U=0.2$.  $\langle S_z^2 \rangle$ reaches its ground state value
  at $\beta<10$ at weak and strong coupling and shows a roughly linear
  dependence on $U$.  In the intermediate coupling regime, $\langle
  S_z^2 \rangle$ has an s-shaped form, and continues to evolve as $T$
  is lowered.  Right: The transitional s-shaped structure in $\langle
  S_z^2 \rangle$ is seen to move to weaker coupling as $J/U$ increases
  at fixed $\beta=14$.  The system is half-filled and size is 8x8.}
\label{S2z}
\end{figure}

The possibility of an OSMT in Eq.~1 with $J_\perp=0$ has been explored
in the DMFT study of Ref.~[\onlinecite{costi07}].  
The local moment $\langle S_z^2 \rangle$ was found to
increase rapidly at a critical value of interaction strength
which is a decreasing function of $J/U$.  At the weakest $J/U < 0.05$,
$\langle S_z^2 \rangle$ exhibits kinks indicative of what proves to be
a first order OSMT.  The evolution of the local moment is smoother for
larger $J/U$, as the OSMT becomes second order.  In Fig.~\ref{S2z} we
show the behavior of $\langle S_z^2 \rangle$ in our DQMC calculations.
Consistent with DMFT, there is an interaction strength, which
decreases as $J/U$ increases, for which the local moment changes
rapidly.  Significantly, in the neighborhood of this $U$ value, the
system must be cooled to a lower temperature in order to reach the
ground state, indicative of the competition between states of nearly
degenerate energy which occurs at a phase boundary.

Within mean field theories, the local moment can act as an order
parameter, since when fluctuations are neglected the distinction
between the energy scales and physics associated with local moment
formation and long range magnetic correlations is blurred.  However,
as is well known, the local moment often loses its sharp structure
when spatial fluctuations are included, as is the case with the DQMC
simulations reported here.  We therefore now turn to other
measurements which can signal the OSMT more clearly.  A key conclusion
of our paper is that these quantities demonstrate that the orbitally
selective transition found in DMFT survives.  

\noindent
\underbar{Density of States:} DQMC allows the direct measurement of
the space and imaginary time Green's function and two particle
correlation functions, 
Frequency dependent quantities
can be obtained through a maximum entropy analytic continuation
procedure\cite{gubernatis91} which inverts the integral relation
between $\omega$ and $\tau$.  

\begin{figure}[t]
\epsfig{figure=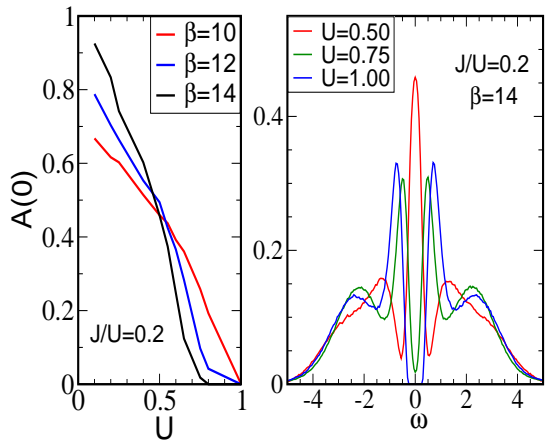,height=8cm,width=7cm,angle=-90,clip}
\caption{(Color online) Left panel: The density of states at the Fermi
  surface $A(\omega=0)$ of the weakly correlated orbital is shown as a
  function of the interaction $U$.  Below $U/t = (U/t)_{\rm crit}
  \approx 0.5 \pm 0.1$, $A(0)$ rises as the temperature is lowered
  ($\beta$ increases), indicative of metallic behavior of the weakly
  interacting band.  At larger $U$, the trend is reversed and $A(0)$
  is suppressed, signaling insulating behavior.
  Right panel: the full frequency dependence of the density of
  states $A(\omega)$ is shown for $J/U=0.2$.
  $A(\omega)$ has a maximum at
  $\omega=0$ for the weakest coupling.  When $U/t=0.75$, a deep
  suppression of $A(\omega)$ is seen at the Fermi surface, and the
  system is fully insulating by the time $U/t=1.0$.
In both panels, 
the lattice is half-filled and has 64 sites.
}
\label{A0vsUJeq0ov5}
\end{figure}

Fig.~\ref{A0vsUJeq0ov5}(left) shows the density of states at the Fermi
surface $A(0)$ for fixed $J/U=0.2$.  We see that as $T \rightarrow 0$,
$A(0)$ is non-zero for $U/t < (U/t)_{c} \approx 0.5 \pm 0.1$.
 Above this critical value, the low temperature limit of $A(0)$ is zero.
$(U/t)_{c}$ lies very close to onset point of $U/t$ at which the local
moment starts exhibiting pronounced temperature dependence,
Fig.~\ref{S2z}(left), as well as to the value $U/t \approx 0.6$ at which
$\langle S_z^2 \rangle$ is changing most rapidly with interaction
strength in Fig.~\ref{S2z}(right).

Fig.~\ref{A0vsUJeq0ov5}(right) exhibits the energy dependence of
$A(\omega)$.  For $U/t$ below the temperature crossing in
Fig.~\ref{A0vsUJeq0ov5}(left), $A(\omega)$ has a maximum at
$\omega=0$, confirming this as a metallic state.  By the time
$U/t=0.75$ this maximum has been replaced by a deep mimimum, almost to
$A(0)=0$.  Indeed, if the temperature were lowered further a full gap
would form, such as is seen for $U/t=1.0$.
Fig.~\ref{A0vsUJeq0ov5}(right) demonstrates that a OSMT occurs in the
Hamiltonian Eq.~1.  The size of the gap $\Delta$ in $A(\omega)$ is
roughly $U$.  However, one typically expects a gap set by $U$ only
deep in the Mott region where $U$ exceeds the bandwidth $W=8t$.  As we
shall discuss further below, we believe that here, instead, the gap
$\Delta$ has a pronounced AF origin and is set by $U m_{\rm af}$ where
$m_{\rm af}$ is the AF order parameter.

\begin{figure}[t]
\epsfig{figure=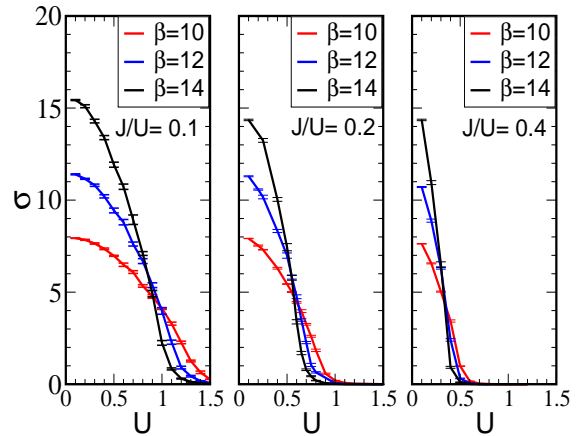,height=8.5cm,width=7cm,angle=-90,clip}
\caption{(Color online) The dc conductivity $\sigma_{\rm dc}$ is shown
  as a function of interaction strength $U/t$ for different values of
  the coupling between the itinerant and localized degrees of freedom.
  Left to right: $J/U=0.1, 0.2, 0.4$.  In all three cases, at weak
  coupling, $\sigma_{\rm dc}$ rises as the temperature $T$ is lowered.
  As for the spectral weight at the Fermi surface
  (Fig.~\ref{A0vsUJeq0ov5}), this indicates the weakly correlated band
  is metallic.  For $U/t$ greater than a critical value, this trend
  with temperature reverses and the weakly correlated band undergoes
  an insulating transition.  $(U/t)_{c}$ decreases as $J/U$ increases.
  The lattice size is 8x8 and the filling $\rho=1$.  }
\label{SigmaN8}
\end{figure}

\noindent
\underbar{Conductivity:} The dc conductivity $\sigma_{\rm dc}$ can be
obtained from the large imaginary time dependence of the
current-current correlation function \cite{trivedi96}.  We show the
results in Fig.~\ref{SigmaN8}.  As with the Fermi surface density of
states, curves for different temperatures $T$ cross when plotted as a
function of $U$.  The intersection demarks a transition from a
metallic phase where $d\sigma_{\rm dc}/dT < 0$ to an insulating phase
with $d\sigma_{\rm dc}/dT > 0$.  The crossing point for $J/U=0.2$ is
consistent with the critical values obtained from $\langle S_z^2
\rangle$ (Fig.~\ref{S2z}) and $A(0)$ (Fig.~\ref{A0vsUJeq0ov5}).

\begin{figure}[t]
\epsfig{figure=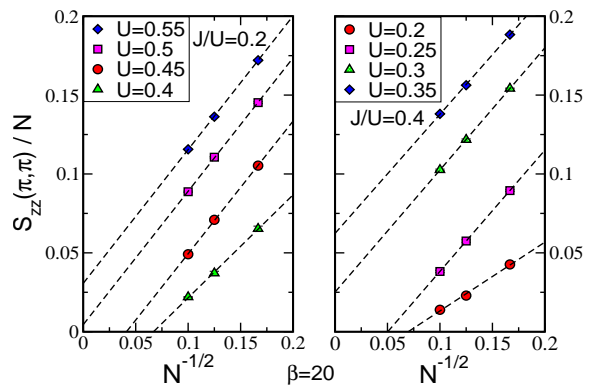,height=8.0cm,width=6.0cm,angle=-90,clip}
\caption{(Color online)
A finite size scaling plot of 
the antiferromagnetic structure factor shows that for $J/U=0.2$ (left
panel), long range order develops for $U$ larger than $U_c \approx
0.5$.  For $J/U=0.4$ (right panel) there is a smaller $U_c \approx 0.3$.
}
\label{AForder}
\end{figure}

\noindent
\underbar{Magnetic Correlations:} The presence of a gap in $A(\omega)$
even when $U$ is an order of magnitude less than the bandwidth
suggests that the insulating behavior does not arise purely from Mott
physics- an energy lowering from avoiding double occupancy exceeding
the cost in kinetic energy.  We now explore the AF correlations which
can give rise to a Slater gap in the spectrum.

Fig.~\ref{AForder} shows the AF structure factor $S_{zz}(\pi,\pi) =
\frac{1}{N} \sum_{i,j} (-1)^{i+j} \langle S_i^z S_j^z \rangle$.  When
long range order is absent, the real space spin correlation $\langle
S_i^z S_j^z \rangle$ decays exponentially. Only sites $j$ within a
correlation length $\xi$ of site $i$ contribute to the sum, and as a
consequence, $S_{zz}(\pi,\pi)$, approaches a lattice size independent
value at large $N$.  $S_{zz}(\pi,\pi) \, / \, N$, shown in
Fig.~\ref{AForder}, therefore vanishes in the thermodynamic limit.  On
the other hand, in an ordered phase, the real space spin correlation
$\langle S_i^z S_j^z \rangle$ is large for all pairs of sites $i,j$.
The structure factor $S_{zz}(\pi,\pi)$, is proportional to $N$, and
$S_{zz}(\pi,\pi) \, / \, N$, shown in Fig.~\ref{AForder}, goes to a
non-zero value.  Huse \cite{huse88} has used spin wave theory to make
this argument more precise, and shown that $S_{zz}(\pi,\pi) \, / \, N
= m_{\rm af}^2/3 + a/L$, where $L=N^{1/2}$ is the linear lattice size
and $m_{\rm af}$ is the AF order parameter.  In Fig.~\ref{AForder} we
see that, at $J/U=0.2$, for small $U$, less than $U_c \approx 0.5$,
$S_{zz}(\pi,\pi) \, / \, N $ goes to zero for large $N$, while for $U$
above this value there is long range order.

These results for long range magnetic order are consistent with the
transition points observed in our early measurements.  For example, in
Fig.~\ref{SigmaN8}, at $J/U=0.2$, when $U=0.5$ (central panel), the
system is metallic.  When $J/U=0.4$ and $U=0.5$ (right panel), in
contrast, the conductivity indicates insulating behavior.  As
expected, $\sigma_{\rm dc}$ goes to zero when there is long range AF
order.  Indeed, the size of the gap $U m_{\rm af}$ which would arise
for electrons of one spin species
moving through a staggered potential due to the other 
also matches well with the values seen in 
Fig.~\ref{A0vsUJeq0ov5}(right).
It is important to note that the interaction term breaks spin
rotation invariance, and hence we have this AF order only in the $z$
direction.  Finite size scaling of the $xy$ AF structure factor
indicates that the associated order parameter vanishes for all
parameter regimes we have studied.

\begin{figure}[t]
\epsfig{figure=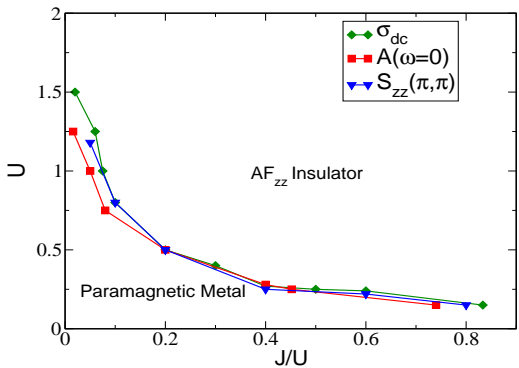,height=7.0cm,width=5.0cm,angle=-90,clip}
\caption{(Color online) The phase diagram of the Hamiltonian Eq.~1, as
  determined by the various quantities measured in our simulation.  At
  weak coupling (small $J/U$ and $U/t$) there is a metallic phase.
	      }
\label{phasediag}
\end{figure}

\noindent
\underbar{Conclusions:} The complete ground state phase diagram in the
$U-J/U$ plane is shown in Fig.~\ref{phasediag}.  The different
observables, $A(\omega=0), \sigma_{\rm dc}$ and $S_{\rm zz}(\pi,\pi)$,
all give (to within our error bars) a common phase boundary which
separates a paramagnetic metallic phase from an insulating
antiferromagnetic one.  Generically, one expects the vertical
($J/U=0$) axis, which corresponds to the usual Hubbard model, to be
insulating above a critical value $U_{\rm c}$ (the Mott transition).
Owing to the (logarithmically) divergent density of states, $U_{\rm
  c}=0$ for a square lattice.  However, this singularity is broken by
a small nonzero $J$.  

Magnetic correlations are known to have an important interplay with
Mott physics both in the single band Hubbard Hamiltonian, and in the
real materials for which it constitutes simple model.  Similarly, in
the single impurity and periodic Anderson Hamiltonians, local singlet
formation and longer range AF order are central phenomena. In this
paper, we have shown that in at least one model, the orbitally
selective Mott transition is accompanied by the formation of a
significant amount of intersite magnetic correlations, and that this
inclusion of spatial fluctuations does not alter the qualitative
physics- a set of itinerant fermions can coexist with a fully
localized band.  As with much earlier work \cite{costi07}, we have
considered a simplified model which does not allow the full Hund's
rule coupling between bands.  Simulations with such a term in place
involve the fermion sign problem and cannot at present be undertaken
at low enough temperatures.

\noindent
\underbar{Acknowledgements:} The work of KB is supported by DOE Grant No. DE-FG02-07ER46423.  RTS
acknowledges funding from DOE-DE-FC0206ER25793.  
G.G.B. is supported by the CNRS
(France) PICS 3659.  We acknowledge very useful help from A. Martin.


\begin{thebibliography}{40}

\bibitem{liebsch04}
A. Liebsch,
Phys. Rev. {\bf B70}, 165103 (2004).

\bibitem{arita05}
R. Arita and K. Held,
Phys. Rev. {\bf B72}, 201102(R) (2005).

\bibitem{inaba06}
K. Inaba and A. Koga,
Phys. Rev. {\bf B73}, 155106 (2006).

\bibitem{liebsch05}
A. Liebsch,
Phys. Rev. Lett. {\bf 95}, 116402 (2005).

\bibitem{orbselective}
M. Ferrero {\it etal.},
Phys. Rev. {\bf B72}, 205126 (2005);
L. De'Medici, A. Georges, and S. Biermann,
Phys. Rev. {\bf B72}, 205124 (2005); 
A. Ru\"egg {\it etal.},
Eur. Phys. J. {\bf B48}, 55 (2005).

\bibitem{DMFT}
Th. Pruschke, M. Jarrell, and J. K. Freericks, Adv. Phys. {\bf 44}, 187
(1995); A. Georges {\it etal.}, 
Rev.  Mod. Phys. {\bf 68}, 13 (1996).

\bibitem{allen82}
J. W. Allen and R. M. Martin, Phys. Rev. Lett. {\bf 49}, 1106, (1982).

\bibitem{held01}
K. Held, A.K. McMahan, and R.T. Scalettar,
Phys.~Rev.~Lett.~{\bf 87}, 276404 (2001).

\bibitem{CaSrRuO}
V.I. Anisimov {\it etal.},
Eur. Phys. J. {\bf B25}, 1434 (2002); 
P.G.J. van Dongen, C. Knecht, and N. Blumer
Phys. Stat. Sol. {\bf (b)243}, 116 (2006); 
M. Neupane {\it etal.},
arXiv:0808.0346.

\bibitem{costi07}
T.A. Costi and A. Liebsch,
Phys. Rev. Lett. {\bf 99}, 236404 (2007).
See also 
N.~Bl\"umer {\it etal},
J. Mag. and Mag. Mat. {\bf 310}, 922 (2007).

\bibitem{liebsch06}
For a further discussion, see
A. Liebsch,
arXiv:cond-mat/0610482 (2006).

\bibitem{imada98}
M. Imada, A. Fujimori, and Y. Tokura,
Rev. Mod. Phys. {\bf 70}, 1039 (1998)
and references cited therein.

\bibitem{nakatsuji0300}
S. Nakatsuji {\it etal.},
Phys. Rev. Lett {\bf 90}, 137202 (2003);
S. Nakatsuji and Y. Maeno, {\it ibid}
{\bf 84}, 2666 (2000).

\bibitem{blankenbecler81}
 R. Blankenbecler, D.J. Scalapino, and R.L. Sugar,
Phys. Rev. {\bf D24}, 2278 (1981).

\bibitem{scalettar91}
R.T.~Scalettar, R.M.~Noack, and R.R.P.~Singh, Phys.~Rev.~{\bf B44},
10502 (1991).

\bibitem{gubernatis91}
J.E. Gubernatis {\it etal.},
Phys. Rev. {\bf B44}, 6011 (1991).

\bibitem{trivedi96}
N.~Trivedi, R.T.~Scalettar, and M.~Randeria,
Phys.~Rev.~{\bf B54}, 3756 (1996).

\bibitem{huse88}
D.A.~Huse,
Phys.~Rev.~{\bf B37}, 2380 (1988).

\end{thebibliography}
\end{document}